\documentclass[12pt]{article}
\usepackage{amssymb}
\usepackage{epsfig}
\usepackage{psfrag}
\usepackage{latexsym}
\usepackage{euscript}
\usepackage{simplewick}
\usepackage{feynmp}
\usepackage{textcomp}
\newcommand{\phit}{\varphi_T}

\newcommand{\phis}{\varphi_S}

\def\gs{\mathrel{
   \rlap{\raise 0.511ex \hbox{$>$}}{\lower 0.511ex \hbox{$\sim$}}}}
\def\ls{\mathrel{
   \rlap{\raise 0.511ex \hbox{$<$}}{\lower 0.511ex \hbox{$\sim$}}}}

\newcommand{\ba}{\begin{array}{c}}
\newcommand{\baz}{\begin{array}{cc}}
\newcommand{\bad}{\begin{array}{ccc}}
\newcommand{\ea}{\end{array}}

\newcommand{\be}{\beta}

\newcommand{\om}{\omega}

\newcommand{\onep}{1^\prime}
\newcommand{\onepp}{1^{\prime\prime}}

\def\beq{\begin{equation}}
\def\eeq{\end{equation}}
\def\bea{\begin{eqnarray}}
\def\eea{\end{eqnarray}}
\def\bet{\begin{tabular}}
\def\eet{\end{tabular}}
\def\bes{\begin{subequations}\bea}
\def\ees{\eea\end{subequations}}
\def\ol{\overline}

\newcommand{\dv}{\partial\hspace{-7pt}\slash}

\newcommand{\D}{D\hspace{-8pt}\slash}
\def\phis{\varphi_S}
\def\phit{\varphi_T}
\def\onep{1'}
\def\onepp{1''}
\def\om{\omega}

\def\be{\begin{equation}}
\def\ee{\end{equation}} 
\def\ol{\overline}
\def\bea{\begin{eqnarray}}
\def\eea{\end{eqnarray}}
\def\nn{\nonumber}

\def\be{\begin{equation}}
\def\ee{\end{equation}}
\def\bc{\begin{center}}
\def\ec{\end{center}}
\def\bea{\begin{eqnarray}}
\def\eea{\end{eqnarray}}

\def\nn{\nonumber}

\catcode`@=11
\def\marginnote#1{}
\newcount\hour
\newcount\minute
\newtoks\amorpm
\hour=\time\divide\hour by60
\minute=\time{\multiply\hour by60 \global\advance\minute by-\hour}
\edef\standardtime{{\ifnum\hour<12 \global\amorpm={am}%
        \else\global\amorpm={pm}\advance\hour by-12 \fi
        \ifnum\hour=0 \hour=12 \fi
        \number\hour:\ifnum\minute<10 0\fi\number\minute\the\amorpm}}
\edef\militarytime{\number\hour:\ifnum\minute<10 0\fi\number\minute}
\def\draftlabel#1{{\@bsphack\if@filesw {\let\thepage\relax
   \xdef\@gtempa{\write\@auxout{\string
      \newlabel{#1}{{\@currentlabel}{\thepage}}}}}\@gtempa
   \if@nobreak \ifvmode\nobreak\fi\fi\fi\@esphack}
        \gdef\@eqnlabel{#1}}
\def\@eqnlabel{}
\def\@vacuum{}
\def\draftmarginnote#1{\marginpar{\raggedright\scriptsize\tt#1}}
\def\draft{\oddsidemargin 0.0truein
        \def\@oddfoot{\sl preliminary draft \hfil
        \rm\thepage\hfil\sl\today\quad\militarytime}
        \let\@evenfoot\@oddfoot \overfullrule 3pt
        \let\label=\draftlabel
        \let\marginnote=\draftmarginnote
   \def\@eqnnum{(\theequation)\rlap{\kern\marginparsep\tt\@eqnlabel}%
\global\let\@eqnlabel\@vacuum}  }
\catcode`@=12

\begin{document}
\begin{titlepage}
\vspace*{-1cm}
\phantom{hep-ph/***}
\hfill{DFPD-10/TH/9}\\
\vskip 2.5cm
\begin{center}
\mathversion{bold}
{\Large\bf Rare muon and tau decays in $A_4$ Models}
\mathversion{normal}
\end{center}
\vskip 0.5  cm
\begin{center}
{\large Ferruccio Feruglio}~\footnote{e-mail address: feruglio@pd.infn.it} and
{\large Alessio Paris}~\footnote{e-mail address: paris@pd.infn.it}
\\
\vskip .2cm
Dipartimento di Fisica `G.~Galilei', Universit\`a di Padova
\\
INFN, Sezione di Padova, Via Marzolo~8, I-35131 Padua, Italy
\end{center}
\vskip 0.7cm
\begin{abstract}
\noindent
We analyze the most general dimension-six effective Lagrangian, invariant under the flavour symmetry $A_4\times Z_3\times U(1)_{FN}$ proposed to reproduce the near tri-bimaximal lepton mixing
observed in neutrino oscillations. The effective Lagrangian includes four-lepton operators that violate the individual lepton numbers in the limit of exact flavor symmetry and allow unsuppressed processes satisfying the rule  
$\Delta L_e \Delta L_\mu \Delta L_\tau=\pm2$.
The most stringent bounds on the strength of the new interactions come from the observed universality of leptonic muon and tau decays, from the agreement
between the Fermi constant measured in the muon decay and that extracted from the $m_W/m_Z$ ratio, and from the limits on the rare decays $ \tau^- \rightarrow \mu^+ e^- e^-$ and $\tau^- \rightarrow e^+ \mu^- \mu^-$.
We also investigate these effects in a specific supersymmetric (SUSY) realization of the flavour symmetry and we find large suppression factors for all the processes allowed by the selection rule.
We explain why this rule is violated in the SUSY context and we provide a complete picture of lepton flavour violation in the SUSY version of $A_4\times Z_3\times U(1)_{FN}$.
\end{abstract}
\end{titlepage}
\setcounter{footnote}{0}
\vskip2truecm
%
%
\section{Introduction}
The discovery of neutrino oscillations has added unexpected features to the flavour puzzle. Two lepton mixing angles are large, one of them being compatible with maximal mixing. 
The third angle is smaller than the other two and it may actually vanish. The central values of the mixing angles are remarkably close
to those predicted by the tri-bimaximal (TB) mixing pattern \cite{TB}, whose mixing matrix has a very symmetric form. There is still a certain amount of experimental uncertainty, 
particularly on the smallest angle and on the atmospheric one, and at the moment we do not know whether in the future the TB mixing will remain a good approximation of the data or not.
Two attitudes are possible. Either the TB pattern is an illusion due to our current poor experimental resolution, lacking in dynamical meaning, or
such a simple pattern is hinting at some important features of an underlying theory of flavour. 

By inspecting the symmetry properties of charged lepton and neutrino mass matrices, it became apparent that TB mixing can arise from the spontaneous breaking of the flavour symmetry group 
$A_4$ \cite{A4}. A simple and economic realization of this idea  \cite{OurTriBi,Altarelli:2005yx,AM} adopts the symmetry group $A_4\times Z_3\times U(1)_{FN}$, whose breaking is
driven by the vacuum alignment of a set of flavon fields. The vacuum expectation values of these fields, in units of some fundamental scale $\Lambda$, provide small expansion parameters $t$ and $u$ in terms
of which all relevant physical quantities can be expressed. An exact TB mixing is only expected in the limit of infinitely small $t$ and $u$ and in the real world corrections of order $u\approx 0.01$ are expected.
Explicit models with such a symmetry breaking pattern have been built, where the vacuum alignment is completely natural, being the result of the minimization of the
most general scalar potential invariant under $A_4\times Z_3\times U(1)_{FN}$, in a finite portion of the parameter space. The charged lepton mass hierarchy arises naturally,
a la Froggatt-Nielsen \cite{FN}, from the breaking of the $U(1)_{FN}$ component induced by $t$. Neutrino masses are constrained in this picture and a number of experimental tests
are already possible within the neutrino sector alone \cite{review}. Nevertheless it would be highly desirable to test these models beyond the neutrino sector.

This possibility can be offered by processes with lepton
flavour violation (LFV) expected, at some level, for massive and non-trivially mixed neutrinos. At low-energy LFV is described by dimension six operators,
suppressed by two powers of a new physics scale $M$, which could be much smaller than the fundamental scale $\Lambda$, thus allowing observable effects.
In models based on flavour symmetries rates of LFV transitions usually receive a double suppression: one from $1/M^4$ and one from the parameters that break the flavour symmetry, in our case $t$ and $u$.
In a previous set of papers \cite{Feruglio:2008ht,Feruglio:2009iu,Feruglio:2009hu} radiative decays of the charged leptons have been analyzed in models invariant under $A_4\times Z_3\times U(1)_{FN}$, finding
a generic suppression $u^2/M^4$ for the rates, that can become more severe in a supersymmetric (SUSY) realization of $A_4\times Z_3\times U(1)_{FN}$. For instance, in this last case the decay rate
of $\mu\to e \gamma$ can scale as $t^2 u^2/M^4$, leaving room to a relatively light scale of new physics $M$. 

In this paper we complete the analysis of LFV in $A_4\times Z_3\times U(1)_{FN}$ symmetric models, by building the most general dimension six effective Lagrangian allowed by the symmetry. 
Contrary to the expectations of many models based on flavour symmetries, such effective Lagrangian contains four-lepton operators that break the conservation of the individual lepton numbers, 
while being fully invariant under the flavour symmetry \cite{Feruglio:2008ht}. These operators leads to LFV transitions whose rates are suppressed only by $1/M^4$, which can result in strong bounds on the  scale $M$. 
We carefully analyze all such transitions, that satisfy the selection rule $\Delta L_e \Delta L_\mu \Delta L_\tau=\pm2$. We separate our discussion according to the type of leptons involved in the transition: four neutrinos,
two neutrinos plus two charged leptons, and four charged leptons. We find that the strength of the new LFV operators are most severely bound by the observed universality of leptonic muon and tau decays, from the agreement
between the Fermi constant measured in the muon decay and that extracted from the $m_W/m_Z$ ratio, and from the limits on the rare decays $ \tau^- \rightarrow \mu^+ e^- e^-$ and $\tau^- \rightarrow e^+ \mu^- \mu^-$.
The present experimental limits on the branching ratios of these two tau decays push the scale $M$ above 10 TeV.

We also analyze a specific SUSY realization of $A_4\times Z_3\times U(1)_{FN}$, motivated by several considerations: it offers a natural solution to the required vacuum alignment,
it might be required to realize the embedding of the model in a grand unified theory and it provides a natural framework for a relatively small scale of new physics $M$, related to the SUSY breaking scale. 
In this SUSY model the four-lepton LFV operators arise from box diagrams, with neutralinos, charginos and sleptons circulating in the loop. 
For all processes that were allowed by the selection rule $\Delta L_e \Delta L_\mu \Delta L_\tau=\pm2$, we find rates suppressed at least by eight powers of the symmetry breaking parameters $t$ and/or $u$,
at variance with the result of the model-independent analysis. We provide a detail explanation for this singular behavior and we reconsider all possible LFV transitions in the SUSY model.
We end up with a complete picture of most relevant LFV processes for the model at hand.

Our work is organized as follows: after a short review of the main features of $A_4$, we will build the most general four-lepton effective Lagrangian invariant under $A_4\times Z_3\times U(1)_{FN}$
and allowing for LFV. Subsequently we will discuss the bounds on the Lagrangian from the available data. Finally we will analyze the specific SUSY version of $A_4\times Z_3\times U(1)_{FN}$
and we will go to the conclusion.
%
%
\section{Brief review of the model}
We will consider a model based on the flavour symmetry group
\begin{equation}
G_f=A_4\times Z_3\times U(1)_{FN}
\label{flavgroup}
\end{equation}
that has been especially tailored to approximately reproduce in a simple and economic way the TB mixing scheme \cite{OurTriBi,Altarelli:2005yx}.
The three factors in $G_f$ play different roles. The spontaneous breaking of the first one, $A_4$, is directly responsible for the TB mixing.   
The $Z_3$ factor is a discrete version of the total lepton number and is needed in order to avoid large mixing effects
among the flavons that give masses to the charged leptons and those giving masses to neutrinos. Finally, $U(1)_{FN}$ is
responsible for the hierarchy among charged fermion masses. The group $A_4$ has twelve elements, three unidimensional representations $1,1',1''$ and a three dimensional representation $3$. 
For our purposes, it will be useful to report explicitly the tensor products of the group representations:
\bea
1\otimes (1,1',1'',3) &=& (1,1',1'',3)\nn \\
1'\otimes 1'&=&1'' \nn \\
1''\otimes 1''&=&1' \nn\\
1'\otimes 1'' &=& 1 \nn \\
1,1',1'' \otimes 3 &=& 3 \nn\\
3\otimes 3 &=& 1 \oplus 1' \oplus 1'' \oplus 3_S \oplus 3_A~~~,
\eea
where $3_S$ and $3_A$ denote symmetric and antisymmetric combinations, respectively.
Consider two general triplets 
\be
a= (a_1,a_2,a_3)~~~,\quad b=(b_1,b_2,b_3)~~~,
\ee
they can be combined into one dimensional representations (we work in the basis of ref. \cite{Altarelli:2005yx}): 
\bea
1\equiv (ab) &=& (a_1b_1+a_2b_2+a_3 b_3)\nn \\
1'\equiv (ab)' &=& (a_3b_3+a_1b_2+a_2 b_1)\nn \\
1''\equiv (ab)'' &=& (a_2b_2+a_1b_3+a_3 b_1)~~~,
\eea
or into two new triplets, a symmetric one and an antisymmetric one.
\bea
3_S\equiv (ab)_S &=& \frac{1}{3}(2a_1b_1-a_2b_3-a_3b_2,2a_3b_3-a_1b_2-a_2b_1,2a_2b_2-a_1b_3-a_3b_1)\nn \\
3_A\equiv (ab)_A &=& \frac{1}{2}(a_2b_3-a_3b_2,a_1b_2-a_2b_1,a_3b_1-a_1b_3)~~~.
\eea
Moreover, given the singlets $c$, $c'$ and $c''$ transforming as $1$, $1'$ and $1''$, the products $ac$, $ac'$ and $ac''$ are triplets, whose explicite form is $(a_1c,a_2c,a_3c)$,$(a_3c',a_1c',a_2c')$ and $(a_2c'',a_3c'',a_1c'')$.
Finally, note that the conjugate $a^*$ does not transform as a triplet; instead, $(a_1^*,a_3^*,a_2^*)$ does.

The flavour symmetry breaking sector of the model includes the scalar fields $\varphi_T$, $\varphi_S$, $\xi$ and $\theta$. The transformation properties of the lepton
fields $L$, $E_e$, $E_\mu$, $E_\tau$, of the electroweak scalar doublet $H$ and of the flavon fields have been recalled in table 1.
In our notation $L\equiv (L_e,L_\mu,L_\tau)$ are left-handed SU(2) doublets with hypercharge Y=-1/2, while $E_e$, $E_\mu$ and $E_\tau$
are right-handed  SU(2) singlets with hypercharge $Y=-1$. Chirality projectors are understood. The following pattern of VEVs for the flavon fields
\begin{eqnarray}
\frac{\langle\varphi_T\rangle}{\Lambda}&=&(u,0,0)+O(u^2)\nonumber\\
\frac{\langle\varphi_S\rangle}{\Lambda}&=& c_b(u,u,u)+O(u^2)\nonumber\\
\frac{\langle\xi\rangle}{\Lambda}&=&c_a u+O(u^2)\nonumber\\
\frac{\langle\theta\rangle}{\Lambda}&=&t
\label{vevs}
\end{eqnarray}
where $u$ and $t$ are the small, real, symmetry breaking parameters of the theory, guarantees that the lepton mixing
is approximately TB. The parameters $c_{a,b}$ are pure numbers of order one and $\Lambda$ is the cutoff of the theory. 
It is possible to achieve this pattern of VEVs in a natural way, as the result of the minimization of the scalar potential of the theory \cite{OurTriBi,Altarelli:2005yx,deMedeirosVarzielas:2005qg}.
\begin{table}[!ht] 
\centering
                \begin{math}
                \begin{array}{|c||c|c|c|c||c|c|c|c|c|}
                    \hline
                    &&&&&&&&& \\[-9pt]
                    \tt{Field} & L & E_e & E_\mu & E_\tau & H & \phit & \phis & \xi & \theta \\[10pt]
                    \hline
                    &&&&&&&&&\\[-9pt]
                    A_4 & 3 & 1 & \onep & \onepp & 1 & 3 & 3 & 1 &  1 \\[3pt]
                    \hline
                    &&&&&&&&&\\[-9pt]
                    Z_3 & \om & \om & \om & \om  & 1 & 1& \om & \om  & 1 \\[3pt]
                    \hline
                    &&&&&&&&&\\[-9pt]
                    U(1)_{FN} & 0 & -2 & -1 & 0  & 0 & 0 & 0 & 0 & -1  \\[3pt]
                    \hline
                \end{array}
               \end{math} 
            \caption{ The transformation rules of the fields under the
            symmetries associated with the groups \rm{$A_4$}, \rm{$Z_3$} and
            \rm{$U(1)_{FN}$}.}
            \end{table}
\vskip 0.2cm
At the leading order, neglecting the $O(u^2)$ contributions, the mass matrix for
the charged leptons is diagonal with the relative hierarchy described by the parameter $t$. 
To reproduce the correct hierarchy we need
\begin{equation} t \approx 0.05~~~.
\label{tbound}
\end{equation}
In the same approximation, the neutrino mass matrix is diagonalized by the TB mixing:
\begin{equation}
U_{TB}=\left(
\begin{array}{ccc}
\sqrt{2/3}& 1/\sqrt{3}& 0\\
-1/\sqrt{6}& 1/\sqrt{3}& -1/\sqrt{2}\\
-1/\sqrt{6}& 1/\sqrt{3}& +1/\sqrt{2}
\end{array}
\right)~~~.
\label{UTB}
\end{equation}
The $O(u^2)$ contributions in eqs. (\ref{vevs}) and the NLO contribution to the Yukawa couplings give rise
to corrections to the TB mixing of relative order $u$. The symmetry breaking parameter $u$ should approximately lie in the range
\begin{equation}
0.005< u < 0.05~~~,
\label{ubound}
\end{equation}
the lower bound coming from the requirement that the Yukawa coupling of the $\tau$ does not exceed $4 \pi$, and the upper bound coming from the
requirement that the higher order corrections, so far neglected, do not modify too much the leading TB mixing.
The inclusion of higher order corrections modifies all mixing angles by quantities of relative order $u$ and in order to keep
the agreement between the predicted and measured values of the solar angle within few degrees, $u$ should not exceed 
approximately 0.05. The unknown angle $\theta_{13}$ is expected to be of order $u$,
not far from the future aimed for experimental sensitivity.
Such a framework can also be extended to the quark sector \cite{quarks,Altarelli:2005yx}. Constraints from baryogenesis have been discussed in ref. \cite{leptogenesis}.

The fields of table 1 and their transformation properties are common to a generic class of models, differing from each other by the specific
mechanism leading to the desired vacuum alignment, eq. (\ref{vevs}), and by additional heavy degrees of freedom. One such model has been realized in the SUSY 
framework \cite{Altarelli:2005yx}, where the special properties of the scalar potential in the SUSY limit are helpful in obtaining the correct vacuum structure.
To construct a general low-energy effective Lagrangian depending on lepton fields we only need the information contained in table 1 and we do not need to specify any particular model,
but in the final part of our work we will make contact with the SUSY realization of ref. \cite{Altarelli:2005yx}. 
%
%
\section{Classification of four-lepton operators}
A complete basis of four-lepton operators, invariant under the SU(2)$\times$ U(1) gauge symmetry, have been introduced by Buchm\"uller and Wyler in \cite{Buchmuller:1985jz}. 
Up to flavour combination, it consists of four independent dimension-six operators:
\bea
 (\mathcal{O}_{LE})^{\beta \delta}_{\alpha \gamma}
 &=&
(\bar{L}^{\beta} E_{\gamma}) (\bar{E}^{\delta} L_{\alpha})  \, , \label{equ:ole} \\
(\mathcal{O}_{LL}^{{\bf 1}})^{\beta \delta}_{\alpha \gamma}
 &=&
 (\bar{L}^{\beta} \gamma^{\rho} L_{\alpha})
 (\bar{L}^{\delta} \gamma_{\rho} L_{\gamma}) \, ,
\label{equ:oll1}
 \\
 (\mathcal{O}_{LL}^{{\bf 3}})^{\beta \delta}_{\alpha \gamma}
 &=&
 (\bar{L}^{\beta} \gamma^{\rho} \vec{\tau} L_{\alpha})
 (\bar{L}^{\delta} \gamma_{\rho} \vec{\tau} L_{\gamma}) \, ,
\label{equ:oll3} \\
(\mathcal{O}_{EE})^{\beta \delta}_{\alpha \gamma} &=& 
 (\bar{E}^\beta \gamma^{\rho}E_\alpha )(\bar{E}^\delta \gamma_{\rho} E_\gamma)  \, .
\label{operators}
\eea
We are using a four-component spinor notation and $\vec{\tau}$ denotes the Pauli matrices acting on SU(2) indices. Greek letters specify the flavour content. Their possible values are $e$, $\mu$ and $\tau$. The effective Lagrangian, invariant under SU(2)$\times$ U(1) gauge transformations,
and depending on $L_\alpha$ and $E_\beta$ is given by:
\bea
\mathcal{L }_{eff} =&-&2\sqrt{2} G_F  \left[
(\mathcal{\varepsilon}_{LE})^{\alpha \gamma}_{\beta \delta} (\mathcal{O}_{LE})^{\beta \delta}_{\alpha \gamma}+
(\mathcal{\varepsilon}_{LL}^{{\bf 1}})^{\alpha \gamma}_{\beta \delta} (\mathcal{O}_{LL}^{{\bf 1}})^{\beta \delta}_{\alpha \gamma}\right.\nn\\
&+&\left.(\mathcal{\varepsilon}_{LL}^{{\bf 3}})^{\alpha \gamma}_{\beta \delta} (\mathcal{O}_{LL}^{{\bf 3}})^{\beta \delta}_{\alpha \gamma}+
(\mathcal{\varepsilon}_{EE})^{\alpha \gamma}_{\beta \delta} (\mathcal{O}_{EE})^{\beta \delta}_{\alpha \gamma} \right]+...
\label{leff}
\eea
We have normalized the interaction strength to the Fermi constant $G_F$, which here we define by the relation
\footnote{It is useful to keep $G_F$ distinguished from $G_\mu$, the constant extracted from muon decay.}
\be
R\equiv\frac{m_W^2}{m_Z^2}=\frac{1}{2}+\sqrt{\frac{1}{4}-\frac{\pi\alpha_{em}(m_Z^2)}{\sqrt{2} G_F m_Z^2 (1-\Delta r)} }
\label{gf}
\ee
where $m_{W,Z}$ are the electroweak gauge boson masses, $\alpha_{em}(m_Z^2)$ is the running QED coupling constant evaluated at the $m_Z$ scale, and $\Delta r$
is the relevant SM radiative correction. This relation defines the constant $G_F$ in terms of the experimental value of the $W$ and $Z$ boson masses.
The Lagrangian ${\cal L}_{eff}$ is hermitian under the following conditions:
\bea
(\mathcal{\varepsilon}_{LE})^{\alpha \gamma}_{\beta \delta}&=&(\mathcal{\varepsilon}_{LE})^{\beta \delta}_{\alpha \gamma}\nn\\
(\mathcal{\varepsilon}_{LL}^{{\bf 1,3}})^{\alpha \gamma}_{\beta \delta}&=&(\mathcal{\varepsilon}_{LL}^{{\bf 1,3}})^{\beta \delta}_{\alpha \gamma}\nn\\
(\mathcal{\varepsilon}_{EE})^{\alpha \gamma}_{\beta \delta}&=&(\mathcal{\varepsilon}_{EE})^{\beta \delta}_{\alpha \gamma}~~~.
\eea
Dots in (\ref{leff}) stand for higher order operators. The most general dimension-six effective Lagrangian depending on lepton fields also
include other operators that we mention for completeness. They fall into two classes. The first one includes operators of dipole type, describing leptonic electric and magnetic dipole moments and flavour changing radiative
transitions such as $\mu\to e \gamma$, $\tau\to \mu\gamma$ and $\tau\to e \gamma$:
\bea
(\mathcal{O}_{dip}^B)^\beta_\alpha&=&\bar{E}^\beta \sigma^{\mu\nu} B_{\mu\nu} H^\dagger L_\alpha\nn\\
(\mathcal{O}_{dip}^W)^\beta_\alpha&=&\bar{E}^\beta \sigma^{\mu\nu} \vec{\tau}~ \vec{W}_{\mu\nu} H^\dagger L_\alpha~~~.
\eea
Their effect in the model under consideration has been analyzed in refs. \cite{Feruglio:2008ht,Feruglio:2009iu,Feruglio:2009hu}.
The second class includes operators bilinear in the lepton fields containing
a derivative and a double insertion of the Higgs multiplet $H$:       
\bea
(\mathcal{O}_{LH}^{{\bf 1}})^{\beta}_{\alpha}
&=&
\left(
 \bar{L}^{\beta} H
\right)
{\rm i} \dv
(H^{\dagger} L_{\alpha})
\nonumber\\
(\mathcal{O}_{LH}^{{\bf 3}})^{\beta}_{\alpha}
&=&
\left(
 \bar{L}^{\beta} \vec{\tau} H
\right)
{\rm i} \D
(H^{\dagger} \vec{\tau} L_{\alpha})
\nonumber\\
(\mathcal{O}_{EH})^{\beta}_{\alpha}
&=&
 \left(
 H^{\dagger}
 {\rm i} D^{\rho}
 H
 \right)
\left(
 \bar{E}^{\beta} \gamma_{\rho} E_{\alpha}
 \right)~~~,
\eea
where $D$ denotes the SM covariant derivative.
They modify kinetic terms of neutrinos and charged leptons and lead to deviations in the neutral and charged
leptonic currents as well as to deviation from unitarity in the leptonic mixing matrix.
Their effects have been discussed in general in refs. \cite{Broncano:2002rw,Broncano:2003fq,Abada:2007ux}. In the model under consideration
the leading contribution to this second class of operators is flavour conserving and will not be further discussed here.

The operators $(\mathcal{O}_{LL}^{{\bf 1}})^{\beta \delta}_{\alpha \gamma}$ and $(\mathcal{O}_{LL}^{{\bf 3}})^{\beta \delta}_{\alpha \gamma}$ only differs by the contraction
of the SU(2) indices. In terms of SU(2) components, they are given by:
\be
(\mathcal{O}_{LL}^{{\bf 1}})^{\beta \delta}_{\alpha \gamma}=
\left(\bar{\nu}^\beta \gamma^\rho \nu_\alpha+\bar{e}^\beta \gamma^\rho e_\alpha\right)
\left(\bar{\nu}^\delta \gamma_\rho \nu_\gamma+\bar{e}^\delta \gamma_\rho e_\gamma\right)
\ee
\bea
(\mathcal{O}_{LL}^{{\bf 3}})^{\beta \delta}_{\alpha \gamma}&=&
\left(\bar{\nu}^\beta \gamma^\rho e_\alpha+\bar{e}^\beta \gamma^\rho \nu_\alpha\right)
\left(\bar{\nu}^\delta \gamma_\rho e_\gamma+\bar{e}^\delta \gamma_\rho \nu_\gamma\right)\nn\\&-&
\left(\bar{\nu}^\beta \gamma^\rho e_\alpha-\bar{e}^\beta \gamma^\rho \nu_\alpha\right)
\left(\bar{\nu}^\delta \gamma_\rho e_\gamma-\bar{e}^\delta \gamma_\rho \nu_\gamma\right)\nn\\&+&
\left(\bar{\nu}^\beta \gamma^\rho \nu_\alpha-\bar{e}^\beta \gamma^\rho e_\alpha\right)
\left(\bar{\nu}^\delta \gamma_\rho \nu_\gamma-\bar{e}^\delta \gamma_\rho e_\gamma\right)~~~.
\eea
The flavour symmetry $G_f$ imposes some restrictions on the coefficients of $\mathcal{L }_{eff}$. 
In a low-energy approximation, below the scale of flavour symmetry breaking, the four-lepton operators of the model originate either from genuine dimension-six operators invariant under $G_f$
or from higher-dimensional $G_f$-invariant operators involving the insertions of the flavon multiplets. After the breaking of $G_f$ by the VEVs
in eq. (\ref{vevs}), the latters become four-lepton operators proportional to some positive power of the symmetry breaking parameters $t$ and/or $u$.  
As a consequence, the coefficients $\mathcal{\varepsilon}^{\alpha \gamma}_{\beta \delta}\equiv 
\{(\mathcal{\varepsilon}_{LE})^{\alpha \gamma}_{\beta \delta},(\mathcal{\varepsilon}_{LL}^{{\bf 1}})^{\alpha \gamma}_{\beta \delta}, (\mathcal{\varepsilon}_{LL}^{{\bf 1}})^{\alpha \gamma}_{\beta \delta}, (\mathcal{\varepsilon}_{EE})^{\alpha \gamma}_{\beta \delta}\}$
can be expanded in powers of the symmetry breaking parameters $t$ and $u$. 
\be
\mathcal{\varepsilon}^{\alpha \gamma}_{\beta \delta}=(\mathcal{\varepsilon}^{(0)})^{\alpha \gamma}_{\beta \delta}+(\mathcal{\varepsilon}^{(1,0)})^{\alpha \gamma}_{\beta \delta}~t+(\mathcal{\varepsilon}^{(0,1)})^{\alpha \gamma}_{\beta \delta}~u+...
\ee
Given the smallness of these parameters, here we will focus on the leading terms  $(\mathcal{\varepsilon}^{(0)})^{\alpha \gamma}_{\beta \delta}$, that is on the operators that do not vanish
when the symmetry breaking effects are neglected. In particular, there are four-lepton operators that violate flavour while being invariant under $G_f$ and not suppressed by powers of $t$ and/or $u$.
We will classify them and we will study their effects. We find:
\bea
\mathcal{L }_{eff} =&-&2\sqrt{2} G_F  \alpha\left(\bar{E}^\tau \gamma^\rho E_\mu ~\bar{E}^e \gamma_\rho E_\mu+\bar{E}^\mu \gamma^\rho ~E_\tau \bar{E}^\mu \gamma_\rho E_e\right)\nn\\
& &\nn\\
&-&2\sqrt{2} G_F \alpha_1\left( \bar{L}^e\gamma^\rho L_e~\bar{L}^e\gamma_\rho L_e+4~\bar{L}^\mu\gamma^\rho L_\mu~\bar{L}^\tau\gamma_\rho L_\tau+\right.\nn\\
& &~~~~~~~~~~~~~~\left. 2~\bar{L}^\mu\gamma^\rho L_e~\bar{L}^\tau\gamma_\rho L_e+2~\bar{L}^e\gamma^\rho L_\mu~\bar{L}^e\gamma_\rho L_\tau\right)\nn\\
&-&2\sqrt{2} G_F \beta_1\left( \bar{L}^\mu\gamma^\rho L_\mu~\bar{L}^\mu\gamma_\rho L_\mu+4~\bar{L}^\tau\gamma^\rho L_\tau~\bar{L}^e\gamma_\rho L_e+\right.\nn\\
& &~~~~~~~~~~~~~~\left. 2~\bar{L}^\tau\gamma^\rho L_\mu~\bar{L}^e\gamma_\rho L_\mu+2~\bar{L}^\mu\gamma^\rho L_\tau~\bar{L}^\mu\gamma_\rho L_e\right)\nn\\
&-&2\sqrt{2} G_F \gamma_1\left( \bar{L}^\tau\gamma^\rho L_\tau~\bar{L}^\tau\gamma_\rho L_\tau+4~\bar{L}^e\gamma^\rho L_e~\bar{L}^\mu\gamma_\rho L_\mu+\right.\nn\\
& &~~~~~~~~~~~~~~\left. 2~\bar{L}^e\gamma^\rho L_\tau~\bar{L}^\mu\gamma_\rho L_\tau+2~\bar{L}^\tau\gamma^\rho L_e~\bar{L}^\tau\gamma_\rho L_\mu\right)\nn\\
&&\nn\\
&-&2\sqrt{2} G_F \alpha_3\left( \bar{L}^e\gamma^\rho\vec{\tau} L_e~\bar{L}^e\gamma_\rho\vec{\tau} L_e+4~\bar{L}^\mu\gamma^\rho\vec{\tau} L_\mu~\bar{L}^\tau\gamma_\rho\vec{\tau} L_\tau+\right.\nn\\
& &~~~~~~~~~~~~~~\left. 2~\bar{L}^\mu\gamma^\rho\vec{\tau} L_e~\bar{L}^\tau\gamma_\rho\vec{\tau} L_e+2~\bar{L}^e\gamma^\rho\vec{\tau} L_\mu~\bar{L}^e\gamma_\rho\vec{\tau} L_\tau\right)\nn\\
&-&2\sqrt{2} G_F \beta_3\left( \bar{L}^\mu\gamma^\rho\vec{\tau} L_\mu~\bar{L}^\mu\gamma_\rho\vec{\tau} L_\mu+4~\bar{L}^\tau\gamma^\rho\vec{\tau} L_\tau~\bar{L}^e\gamma_\rho\vec{\tau} L_e+\right.\nn\\
& &~~~~~~~~~~~~~~\left. 2~\bar{L}^\tau\gamma^\rho\vec{\tau} L_\mu~\bar{L}^e\gamma_\rho\vec{\tau} L_\mu+2~\bar{L}^\mu\gamma^\rho\vec{\tau} L_\tau~\bar{L}^\mu\gamma_\rho\vec{\tau} L_e\right)\nn\\
&-&2\sqrt{2} G_F \gamma_3\left( \bar{L}^\tau\gamma^\rho\vec{\tau} L_\tau~\bar{L}^\tau\gamma_\rho\vec{\tau} L_\tau+4~\bar{L}^e\gamma^\rho\vec{\tau} L_e~\bar{L}^\mu\gamma_\rho\vec{\tau} L_\mu+\right.\nn\\
& &~~~~~~~~~~~~~~\left. 2~\bar{L}^e\gamma^\rho\vec{\tau} L_\tau~\bar{L}^\mu\gamma_\rho\vec{\tau} L_\tau+2~\bar{L}^\tau\gamma^\rho\vec{\tau} L_e~\bar{L}^\tau\gamma_\rho\vec{\tau} L_\mu\right)+...
\label{effa4}
\eea
where dots stands for operators that do not violate lepton flavour. For instance, there are three independent operators of the type $(\mathcal{O}_{LE})^{\beta \delta}_{\alpha \gamma}$, but they are all flavour conserving.
Notice that among the operators of the type $(\mathcal{O}_{EE})^{\beta \delta}_{\alpha \gamma}$ only one of them is flavour violating.
We found three independent operators of the type $(\mathcal{O}_{LL}^{{\bf 1}})^{\beta \delta}_{\alpha \gamma}$ that violate lepton flavour, and we call the corresponding coefficients $\alpha_1$, $\beta_1$ and $\gamma_1$.
Similarly, there are three independent flavor-violating operators of the type $(\mathcal{O}_{LL}^{{\bf 3}})^{\beta \delta}_{\alpha \gamma}$, entering the Lagrangian with weights given by the coefficients $\alpha_3$, $\beta_3$ and $\gamma_3$.
All other operators in ${\cal L}_{eff}$ either conserve flavour or are suppressed by some power of the symmetry breaking parameters $t$ and $u$. From the Lagrangian (\ref{effa4}) it is clear that symmetry restricts the allowed operators to a class satisfying the selection rule $\Delta L_e \Delta L_\mu \Delta L_\tau=0,\pm2$. By expanding the Lagrangian (\ref{effa4}) in neutrino and charged lepton components we get:
\be
\mathcal{L }_{eff} =\mathcal{L }_{4\nu} +\mathcal{L }_{decay} +\mathcal{L }_{NSI} +\mathcal{L }_{ch} 
\ee
where 
\bea
\mathcal{L }_{4\nu} &=&-2\sqrt{2}G_F\times\nn\\
&&\left\{(\alpha_1+\alpha_3) ~\bar{\nu}_e \gamma^\rho \nu_e ~\bar{\nu}_e \gamma_\rho \nu_e+(\beta_1+\beta_3)~\bar{\nu}_\mu \gamma^\rho \nu_\mu ~\bar{\nu}_\mu \gamma_\rho \nu_\mu+(\gamma_1+\gamma_3)\bar{\nu}_\tau \gamma^\rho \nu_\tau ~\bar{\nu}_\tau \gamma_\rho \nu_\tau+\right.\nn\\
&&\left.4\left[(\gamma_1+\gamma_3) ~\bar{\nu}_e \gamma^\rho \nu_e ~\bar{\nu}_\mu \gamma_\rho \nu_\mu+(\alpha_1+\alpha_3)~\bar{\nu}_\mu \gamma^\rho \nu_\mu ~\bar{\nu}_\tau \gamma_\rho \nu_\tau+(\beta_1+\beta_3)\bar{\nu}_\tau \gamma^\rho \nu_\tau ~\bar{\nu}_e \gamma_\rho \nu_e\right]+\right.\nn\\
&&\left.2\left[(\alpha_1+\alpha_3)~\bar{\nu}_\mu \gamma^\rho \nu_e ~\bar{\nu}_\tau \gamma_\rho \nu_e+(\beta_1+\beta_3)~\bar{\nu}_\tau \gamma^\rho \nu_\mu ~\bar{\nu}_e \gamma_\rho \nu_\mu+(\gamma_1+\gamma_3)\bar{\nu}_e \gamma^\rho \nu_\tau ~\bar{\nu}_\mu \gamma_\rho \nu_\tau
\right.\right.\nn\\
&&\left.\left.+h.c.\right]\right\}
\label{l4nu}
\eea
\newpage
\bea
\mathcal{L }_{decay} &=&-2\sqrt{2}G_F\times\nn\\
&&\left[
(1+8\gamma_3)~\bar{e} \gamma^\rho \nu_e~\bar{\nu}_\mu\gamma_\rho \mu+2(\alpha_1+\alpha_3)~\bar{e} \gamma^\rho \nu_\tau~\bar{\nu}_e\gamma_\rho \mu+
2(\beta_1+\beta_3)~\bar{e} \gamma^\rho \nu_\mu~\bar{\nu}_\tau\gamma_\rho \mu+\right.\nn\\
&&\left.(1+8\alpha_3)~\bar{\mu} \gamma^\rho \nu_\mu~\bar{\nu}_\tau\gamma_\rho \tau+2(\beta_1+\beta_3)~\bar{\mu} \gamma^\rho \nu_e~\bar{\nu}_\mu\gamma_\rho \tau+
2(\gamma_1+\gamma_3)~\bar{\mu} \gamma^\rho \nu_\tau~\bar{\nu}_e\gamma_\rho \tau+\right.\nn\\
&&\left.(1+8\beta_3)~\bar{e} \gamma^\rho \nu_e~\bar{\nu}_\tau\gamma_\rho \tau+2(\alpha_1+\alpha_3)~\bar{e} \gamma^\rho \nu_\mu~\bar{\nu}_e\gamma_\rho \tau+
2(\gamma_1+\gamma_3)~\bar{e} \gamma^\rho \nu_\tau~\bar{\nu}_\mu\gamma_\rho \tau\right]\nn\\&&+h.c.
\label{ldecay}
\eea
\bea
\mathcal{L }_{NSI} &=&-2\sqrt{2}G_F\times\nn\\
&&\left\{\left[2(\alpha_1+\alpha_3)~\bar{\nu}_e \gamma^\rho \nu_e+4(\gamma_1-\gamma_3)~\bar{\nu}_\mu \gamma^\rho \nu_\mu+4(\beta_1-\beta_3)~\bar{\nu}_\tau \gamma^\rho \nu_\tau\right]~\bar{e}\gamma_\rho e+\right.\nn\\
&&\left.\left[4(\gamma_1-\gamma_3)~\bar{\nu}_e \gamma^\rho \nu_e+2(\beta_1+\beta_3)~\bar{\nu}_\mu \gamma^\rho \nu_\mu+4(\alpha_1-\alpha_3)~\bar{\nu}_\tau \gamma^\rho \nu_\tau\right]~\bar{\mu}\gamma_\rho \mu+\right.\nn\\
&&\left.\left[4(\beta_1-\beta_3)~\bar{\nu}_e \gamma^\rho \nu_e+4(\alpha_1-\alpha_3)~\bar{\nu}_\mu \gamma^\rho \nu_\mu+2(\gamma_1+\gamma_3)~\bar{\nu}_\tau \gamma^\rho \nu_\tau\right]~\bar{\tau}\gamma_\rho \tau\right\}\nn\\
\label{lnsi}
\eea
\bea
\mathcal{L }_{ch} &=&-2\sqrt{2}G_F\times\nn\\
&&\left\{(\alpha_1+\alpha_3) ~\bar{e} \gamma^\rho e ~\bar{e} \gamma_\rho e+(\beta_1+\beta_3)~\bar{\mu} \gamma^\rho \mu ~\bar{\mu} \gamma_\rho \mu+(\gamma_1+\gamma_3)\bar{\tau} \gamma^\rho \tau ~\bar{\tau} \gamma_\rho \tau+\right.\nn\\
&&\left.4\left[(\gamma_1+\gamma_3) ~\bar{e} \gamma^\rho e ~\bar{\mu} \gamma_\rho \mu+(\alpha_1+\alpha_3)~\bar{\mu} \gamma^\rho \mu ~\bar{\tau} \gamma_\rho \tau+(\beta_1+\beta_3)\bar{\tau} \gamma^\rho \tau ~\bar{e} \gamma_\rho e\right]+\right.\nn\\
&&\left.2\left[(\alpha_1+\alpha_3)~\bar{\mu} \gamma^\rho e ~\bar{\tau} \gamma_\rho e+(\beta_1+\beta_3)~\bar{\tau} \gamma^\rho \mu ~\bar{e} \gamma_\rho \mu+(\gamma_1+\gamma_3)\bar{e} \gamma^\rho \tau ~\bar{\mu} \gamma_\rho \tau
+h.c.\right]+\right.\nn\\
&&\left.\alpha\left[\bar{E}^\tau \gamma^\rho E_\mu ~\bar{E}^e \gamma_\rho E_\mu+h.c.\right]\right\}
\label{lcharged}
\eea
In our notation all above fields are left-handed, except the charged leptons $E_{e,\mu,\tau}$ that are right-handed. 
%
%
\section{Bounds}
We discuss the bounds on the parameters $\varepsilon=\{\alpha,\alpha_{1,3},\beta_{1,3},\gamma_{1,3}\}$, by distinguishing three types of operators: those involving neutrinos only, those involving both neutrinos and charged leptons and finally
those depending on charged leptons only. For simplicity we assume that all the new parameters $\varepsilon$ are real.
%
%
\subsection{Neutrino self-interactions}
Neutrino self-interactions are poorly constrained by present data \cite{bilenky1999}. Extremely mild bounds come from astrophysics. For instance limits can be derived by requiring that the mean free path of neutrinos propagating in the medium between the supernova SN1987A and us is comparable to or larger than the distance to the supernova \cite{kolb1987}. Such bounds would allow self-coupling of a four-fermion interaction larger than the Fermi coupling by many order of magnitudes.
Stronger bounds can be derived from the agreement between the observed invisible decay width of the Z boson and the value predicted by the Standard Model (SM). New neutrino self-interactions of the type considered here
would contribute to both the decay width of the Z into four neutrinos at the tree level \cite{bilenky1993}, and to the decay width of the Z into two neutrinos at the one-loop level \cite{bilenky1994}. These new contributions  remain sufficiently small for values of the couplings $\alpha_{1,3},\beta_{1,3},\gamma_{1,3}$ smaller than approximately $1\div 100$. 
\subsection{Neutrino-charged lepton interactions}
%
%
The new operators containing two neutrinos and two charged leptons lead to several effects. Some of them add to the SM Lagrangian to modify the prediction for the purely leptonic decays
of the charged leptons $\mu$ and $\tau$. The relevant Lagrangian for muon and tau decays is $\mathcal{L }_{decay}$,
from which we can identify the ``Fermi'' constants extracted from the decays $\mu\to e \bar{\nu}\nu$, $\tau\to\mu \bar{\nu} \nu$, $\tau\to e\bar{\nu} \nu$,
that we call $G_\mu$, $G_{\tau\mu}$ and $G_{\tau e}$, respectively. By expanding the results in powers of the real parameters $\varepsilon$, we find:
\bea
G_\mu&=&G_F\left(1+8 \gamma_3+2(\alpha_1+\alpha_3)^2+2(\beta_1+\beta_3)^2+... \right)\nn\\
G_{\tau\mu}&=&G_F\left(1+8 \alpha_3+2(\beta_1+\beta_3)^2+2(\gamma_1+\gamma_3)^2+... \right)\nn\\
G_{\tau e}&=&G_F\left(1+8 \beta_3+2(\gamma_1+\gamma_3)^2+2(\alpha_1+\alpha_3)^2+... \right)
\label{gs}
\eea
where dots stand for higher powers of the parameters. We see that the operators $\mathcal{O}_{LL}^{{\bf 3}}$ lead to amplitudes that interfere with the SM ones, so that
the sensitivity to the corresponding parameters is higher, whereas the operators $\mathcal{O}_{LL}^{{\bf 1}}$ contribute to the decays through non-interfering amplitudes.
The presence of the new operators leads to deviations from universality in weak interactions. We have
\bea
\left(\frac{G_{\tau e}}{G_\mu}
\right)^2
=&1+16 (\beta_3-\gamma_3)+4(\gamma_1^2+49 \gamma_3^2-\beta_1^2+15\beta_3^2+2\gamma_1\gamma_3-2\beta_1\beta_3-64 \gamma_3\beta_3)\nn\\
\left(\frac{G_{\tau \mu}}{G_\mu}
\right)^2
=&1+16 (\alpha_3-\gamma_3)+4(\gamma_1^2+49 \gamma_3^2-\alpha_1^2+15\alpha_3^2+2\gamma_1\gamma_3-2\alpha_1\alpha_3-64 \gamma_3\alpha_3)\nn\\
\left(\frac{G_{\tau \mu}}{G_{\tau e}}
\right)^2
=&1+16 (\alpha_3-\beta_3)+4(\beta_1^2+49 \beta_3^2-\alpha_1^2+15\alpha_3^2+2\beta_1\beta_3-2\alpha_1\alpha_3-64 \beta_3\alpha_3)\nn\\
\eea
These ratios can be directly compared with data using the following relations:
\bea
\left(\frac{G_{\tau e}}{G_\mu}
\right)^2&=&\frac{\tau_\mu}{\tau_\tau}BR(\tau^-\to e^-\bar{\nu}_e\nu_\tau)\left(\frac{m_\mu}{m_\tau}\right)^5\frac{f(m_e^2/m_\mu^2) r_{EW}^\mu}{f(m_e^2/m_\tau^2) r_{EW}^\tau}\nn\\
\left(\frac{G_{\tau \mu}}{G_\mu}
\right)^2&=&\frac{\tau_\mu}{\tau_\tau}BR(\tau^-\to \mu^-\bar{\nu}_\mu\nu_\tau)\left(\frac{m_\mu}{m_\tau}\right)^5\frac{f(m_e^2/m_\mu^2) r_{EW}^\mu}{f(m_\mu^2/m_\tau^2) r_{EW}^\tau}\nn\\
\left(\frac{G_{\tau \mu}}{G_{\tau e}}\right)^2&=&\frac{BR(\tau^-\to \mu^-\bar{\nu}_\mu\nu_\tau)}{BR(\tau^-\to e^-\bar{\nu}_e\nu_\tau)}\frac{f(m_e^2/m_\tau^2) }{f(m_\mu^2/m_\tau^2)}
\eea
where $\tau_{\mu,\tau}$ are the muon and tau lifetimes, $f(x)=1-8x+8x^3-x^4-12 x^2 \log x$ and radiative corrections are those of the SM \cite{MS88}
\be
r_{EW}^l=\left(1+\frac{3}{5}\frac{m_l^2}{m_W^2}\right)\left(1+\frac{\alpha(m_l)}{2\pi}\left(\frac{25}{4}-\pi^2\right)\right)~~~,
\ee
\be
\alpha^{-1}(m_l)=\alpha^{-1}-\frac{2}{3\pi}\log\frac{m_l}{m_e}+\frac{1}{6\pi}~~~.
\ee
From the measured values of $\tau_{\tau}$ and of $BR(\tau^-\to \mu^-\bar{\nu}_\mu\nu_\tau)$, $BR(\tau^-\to e^-\bar{\nu}_e\nu_\tau)$ and their ratio
\be
\tau_\tau=(290.6\pm1.1)\times 10^{-15}~s~~~~~~\cite{pdg08}
\ee
\bea
BR(\tau^-\to e^-\bar{\nu}_\mu\nu_\tau)&=&0.1785\pm0.0005~~~~~~\cite{pdg08}\nn\\
BR(\tau^-\to \mu^-\bar{\nu}_\mu\nu_\tau)&=&0.1736\pm0.0005~~~~~~\cite{pdg08}\nn\\
\frac{BR(\tau^-\to \mu^-\bar{\nu}_\mu\nu_\tau)}{BR(\tau^-\to e^-\bar{\nu}_\mu\nu_\tau)}&=&0.9796\pm0.0040~~~~~~\cite{babar09}
\eea
we get
\be
\left(\frac{G_{\tau e}}{G_\mu}
\right)^2=1.0025\pm0.0047~~~~~~\left(\frac{G_{\tau \mu}}{G_\mu}\right)^2=1.0025\pm 0.0048~~~~~~\left(\frac{G_{\tau \mu}}{G_{\tau e}}\right)^2=1.0072\pm 0.0041
\ee

Assuming values of the parameters $\varepsilon$ roughly of the same order, we see that the deviations from the SM prediction are dominated by
the operators of type $\mathcal{O}_{LL}^{{\bf 3}}$, and we obtain the bounds
\bea
-0.0007&<\beta_3-\gamma_3<0.0010~~~~[3\sigma]\nn\\
-0.0007&<\alpha_3-\gamma_3<0.0010~~~~[3\sigma]\nn\\
-0.0003&<\alpha_3-\beta_3<0.0012~~~~[3\sigma]
\eea
When the operators $\mathcal{O}_{LL}^{{\bf 3}}$ can be neglected, we get a milder bound on the new parameters. For instance,
if $\beta_3=\gamma_3=0$, we have
\be
-0.0033<\gamma_1^2-\beta_1^2<0.0043~~~~[3\sigma]
\ee
If we parametrize the effective Lagrangian in terms of a new mass scale $M$, through the relation
\be
2\sqrt{2} G_F\varepsilon =\frac{1}{M^2}~~~
\ee
we see that an upper bound on $|\varepsilon|$ of order 0.001(0.06) corresponds to a lower bound on $M$ of order 5.5(0.7) TeV.

The comparison between $G_F$ in (\ref{gf}), extracted from the $W$ mass, and $G_\mu$ in (\ref{gs}), obtained from the muon lifetime, leads to additional constraints on the parameters.
It is convenient to express the ratio $R=(m_W^2/m_Z^2)$ in terms of $G_\mu$. By expanding up to second order in the new parameters, we get
\be
R=R_{SM} \left[ 1-\frac{(1-R_{SM})}{(2 R_{SM}-1)}\left( 8\gamma_3+2(\alpha_1+\alpha_3)^2+2(\beta_1+\beta_3)^2 \right)-64\frac{R_{SM}(1-R_{SM})^2}{(2 R_{SM}-1)^3}\gamma_3^2 \right]
\label{rus}
\ee
where we have defined
\be
R_{SM}\equiv\frac{1}{2}+\sqrt{\frac{1}{4}-\frac{\pi\alpha_{em}(m_Z^2)}{\sqrt{2} G_\mu m_Z^2 (1-\Delta r)} }=0.77680(0.77611)
\label{gfsm}
\ee
The numerical values have been obtained from ref. \cite{Ferroglia2002} in the OSII scheme, by using as inputs $m_Z=91.1875$ GeV, $\alpha_s(m_Z)=0.118$, $\Delta\alpha_h^{(5)}=0.02761$, $m_t=173.1$ GeV
and $m_H=115(200)$ GeV. The experimental value of $R$, obtained by combining the PDG averages of the $W$ and $Z$ masses is
\be
R_{\tt exp}=0.77735\pm0.00048
\label{rexp}
\ee
We first compare (\ref{rus}) and (\ref{rexp}) assuming dominance of the $\gamma_3$ parameter. We obtain
\be
-0.0010<\gamma_3<0.0004~~~[3\sigma]~~~.
\ee 
If $\gamma_3$ is negligible, we get
\be
-0.0041<(\alpha_1+\alpha_3)^2+(\beta_1+\beta_3)^2<0.0014~~~[3\sigma]~~~.
\ee 
The bounds are approximately in the same range derived from the universality of the muon and tau decays.

Weaker bound are derived from the non-standard neutrino interactions described by the Lagrangian $\mathcal{L}_{NSI}$ in eq. (\ref{lnsi}). These interactions are usually described in terms of
a perturbation of the weak effective  lagrangian,
\be
\mathcal{L}_{NSI}=-2\sqrt{2}G_F[\epsilon_{\alpha \beta}^{lL,R} (\ol{\nu_\alpha}\gamma^\mu \nu_\beta)(\ol{l}_{L,R} \gamma_\mu l_{L,R})],
\label{nsi}
\ee
where $\epsilon_{\alpha \beta}^{lL,R}$ is a small parameter measuring the size of the deviation and $l=e,\mu,\tau$. The strongest bounds are those on neutrino-electron interactions, $\epsilon_{\alpha\beta}^{eL,R}$.
When $\alpha=\beta=e,\mu$ they are derived from neutrino-electron elastic scattering \cite{nsi}
\bea
-0.07<\epsilon_{ee}^{eL}<0.11&~~~~~~~~~~&-1.0<\epsilon_{ee}^{eR}<0.5~~~~~~~~~~~~~[90\%~~~ CL]\nn\\
-0.025<\epsilon_{\mu\mu}^{eL}<0.03&~~~~~~~~~~&-0.027<\epsilon_{\mu\mu}^{eR}<0.03~~~~~~~~~[90\%~~~ CL]
\label{b12}
\eea
For $\alpha=\beta=\tau$ the limit comes from the $e^+e^-\to \bar{\nu} \nu\gamma$ cross-section measured at LEP II
\be
-0.6<\epsilon_{\tau\tau}^{eL}<0.4~~~~~~~~~~-0.4<\epsilon_{\tau\tau}^{eR}<0.6~~~~~~~~~~~~~[90\%~~~ CL]
\label{b3}
\ee
Similar bounds can be derived on flavour-changing terms $\alpha\ne\beta$ from matter effects in neutrino oscillations, but they are not relevant for the present analysis. Indeed,
by comparing eqs. (\ref{lnsi}) and (\ref{nsi}), we see that in our model, only flavour conserving terms are allowed ($\alpha=\beta$) and we have
\be
\epsilon_{ee}^{eL}=2(\alpha_1+\alpha_3)~~~~~~~~~~
\epsilon_{\mu\mu}^{eL}=4(\gamma_1-\gamma_3)~~~~~~~~~~
\epsilon_{\tau\tau}^{eL}=4(\beta_1-\beta_3)~~~,
\ee 
while right-handed couplings are vanishing. From eqs. (\ref{b12}) and (\ref{b3}) we get the $90\%$ CL bounds
\be
-0.04<(\alpha_1+\alpha_3)<0.06~~~~~~~~~~-0.006<(\gamma_1-\gamma_3)<0.007~~~~~~~~~~-0.15<(\beta_1-\beta_3)<0.1~~~.
\ee
With the exception of the limit on $(\gamma_1-\gamma_3)$, at present these bounds are not competitive with those derived previously from muon and tau decays. 
A future improvement is expected from further analysis of data from KamLAND, SNO, SuperKamiokande and neutrino factories. 
\subsection{Charged lepton interactions}
New interactions among charged leptons are induced by the Lagrangian (\ref{lcharged}). We have seen that LFV transitions obey the selection rule 
$\Delta L_e \Delta L_\mu \Delta L_\tau=\pm2$. This is a very interesting feature of the model under discussion.
Usually SM extensions allowing for low-energy flavour changing four-lepton interactions are severely constrained by the 
experimental limits on the branching ratio $\mu\to e e e$: $BR(\mu\to eee)<1.0\times10^{-12}$. 
In our model this transition is forbidden at the 
lowest order by the selection rule imposed by $A_4$ symmetry. For the same reason also the transitions $\tau \to e e e$ and $\tau \to \mu \mu \mu$ are forbidden at the lowest order.
The above selection rule allows the flavour changing decays $ \tau^- \rightarrow \mu^+ e^- e^-$ and $\tau^- \rightarrow e^+ \mu^- \mu^-$, whose branching ratios have the following upper limits at 90\% CL \cite{pdg08}:
\be
BR(\tau^- \rightarrow \mu^+ e^- e^-) < 2.0 \times 10^{-8}~~~~~~~~~~~~~~~~~~~~BR(\tau^- \rightarrow e^+ \mu^- \mu^-) < 2.3\times 10^{-8}~~~.
\ee
From (\ref{lcharged}), we compute the corresponding decay rates
\bea
\Gamma(\tau^- \rightarrow \mu^+ e^- e^-)&=&\frac{G_F^2 m_\tau^5}{96\pi^3} (\alpha_1+\alpha_3)^2\nn\\
\Gamma(\tau^- \rightarrow e^+ \mu^- \mu^-)&=&\frac{G_F^2 m_\tau^5}{96\pi^3}\left[(\beta_1+\beta_3)^2+\frac{\alpha^2}{4}\right].
\eea 
The total $\tau$ width is approximately given by:
\be
\Gamma=\frac{G_F^2 m_\tau^5}{192\pi^3}\times \frac{1}{BR(\tau^-\to\mu^-\nu_\tau\bar{\nu}_\mu)}\approx \frac{G_F^2 m_\tau^5}{192\pi^3}\times \frac{1}{0.18}~~~,
\ee
and we get the 90\% CL bounds
\bea
\vert \alpha_1+\alpha_3\vert& <&2.4\times 10^{-4}\nn\\
\sqrt{(\beta_1+\beta_3)^2+\frac{\alpha^2}{4}}&<&2.5\times 10^{-4}~~~.
\label{bb}
\eea
These bounds are the most restrictive ones among those discussed so far. The effective Lagrangian (\ref{lcharged}) also describes flavour-conserving four-lepton interactions, such as $e^+ e^- \rightarrow f \bar{f}$  ($f=e,\mu,\tau$),
which have been constrained by the LEP data \cite{lepdata}. For instance, in terms of the effective operator $-2\sqrt{2}G_F [\epsilon_{LL} (\bar{e} \gamma^\rho e) (~\bar{f} \gamma_\rho f)]$, ref.  \cite{lepdata} quotes 
\bea
\epsilon_{LL}^{f\neq e}=0.0168 \pm 0.0133&~~~~~~~~~~&\epsilon_{LL}^{f= e}=-0.0187 \pm 0.0320
\label{blep}
\eea
Using (\ref{lcharged}), we obtain the bounds at $3\sigma$
\bea
-0.006<(\gamma_1+\gamma_3)+(\beta_1+\beta_3)<0.014&~~~~~~&-0.124<(\alpha_1+\alpha_3)<0.086~~~.
\eea
Similar bounds exist on four-lepton flavour-conserving operators with different chirality structure. 
\section{A specific realization}
So far we have analyzed the consequences of the $A_4$ flavour symmetry in a general effective Lagrangian approach, without making reference to any
particular model. The only model-dependent ingredient that we have used is the assignment of the lepton multiplets to representations of the full flavour group
$A_4\times Z_3\times U(1)_{FN}$ given in table 1. It is interesting to investigate some concrete realizations of the flavour symmetry in a specific model, to see if 
the expectations based on the effective Lagrangian approach are fulfilled or not. Perhaps the most significant feature of the effective Lagrangian approach  is
the prediction of the leading order selection rule $\Delta L_e \Delta L_\mu \Delta L_\tau=\pm2$, which implies the dominance of the channels
$ \tau^- \rightarrow \mu^+ e^- e^-$ and $\tau^- \rightarrow e^+ \mu^- \mu^-$ among the flavour-changing transitions.
In this section we consider the supersymmetric realization of $A_4\times Z_3\times U(1)_{FN}$ discussed in ref. \cite{Altarelli:2005yx}. The fields in table 1 are promoted to chiral 
supermultiplets and a $U(1)_R$ symmetry, eventually broken down to the $R$-parity , is introduced to restrict the superpotential.
This realization of $A_4\times Z_3\times U(1)_{FN}$ is particularly relevant since it offers a complete and natural solution of the vacuum alignment problem. 
Namely the specific pattern of VEVs given in eq. (\ref{vevs}) can be reproduced from the minimization of the
scalar potential of the theory in a finite portion of the parameter space, without any fine-tuning of the parameters.
After SUSY breaking four-lepton operators are expected to arise at one-loop from the exchange of sleptons, charginos and neutralinos. 
They are naturally depleted by the effective SUSY breaking scale $m_{SUSY}$, so that, in the absence of other suppressions, our parameters 
$\varepsilon$ are expected to be of order
\be
\varepsilon\approx \frac{1}{16\pi^2}\frac{m_Z^2}{m_{SUSY}^2}
\ee
If $m_{SUSY}\approx 1$ TeV we have $\varepsilon\approx 10^{-4}$ which, as we have seen, is close to the experimental upper bounds on the branching ratios for
$ \tau^- \rightarrow \mu^+ e^- e^-$ and $\tau^- \rightarrow e^+ \mu^- \mu^-$, see eq. (\ref{bb}).
For this reason it is important to proceed to a direct estimate of the rates for lepton flavour violating processes in the SUSY model in order to establish their strength and their relative 
hierarchy. We first focus on the potentially most dangerous transitions: $ \tau^- \rightarrow \mu^+ e^- e^-$ and $\tau^- \rightarrow e^+ \mu^- \mu^-$. At one-loop these transitions
are described by box diagrams alone, since penguing diagrams always require a particle-antiparticle pair in the final state. It is useful to analyze these diagrams in the so-called
super-CKM basis, where gaugino-lepton-slepton vertices are flavour diagonal. Neglecting Higgsino exchange, whose contributions are suppressed by lepton masses,
the only sources of flavour change are the off-diagonal terms of slepton mass matrices $(\tilde{m}^2)_{MN}$, which can be analyzed in the mass insertion approximation, expressed through the parameters 
\be
(\delta_{ij})_{M N}=\frac{(\tilde{m}^2_{ij})_{M N}}{m^2_{SUSY}}
\ee
where $M,N=(L,R)$ are the chiralities. A quick inspection of the relevant box 
diagrams reveals that both transitions require at least two mass insertions: $\delta_{\tau e}$ and $\delta_{\mu e}$ for $ \tau^- \rightarrow \mu^+ e^- e^-$ and $\delta_{\tau \mu}$ and $\delta_{\mu e}$ for
$\tau^- \rightarrow e^+ \mu^- \mu^-$ and the corresponding amplitudes scale as:
\be
{\cal M}( \tau^- \rightarrow \mu^+ e^- e^-)\propto \frac{1}{16\pi^2}\frac{m_Z^2}{m_{SUSY}^2} \times \delta_{\tau e} \delta_{\mu e}
\ee
\be
{\cal M}(\tau^- \rightarrow e^+ \mu^- \mu^-)\propto \frac{1}{16\pi^2}\frac{m_Z^2}{m_{SUSY}^2} \times \delta_{\tau \mu} \delta_{\mu e}
\ee
The most general slepton mass matrices compatible with the $A_4\times Z_3\times U(1)_{FN}$ flavour symmetry in the super-CKM basis
have the following leading order structure \cite{Feruglio:2009iu,Feruglio:2009hu}:
\be
\tilde{m}^2_{eLL}=\tilde{m}^2_{\nu LL}
=\left(
\begin{array}{ccc}
1+O(u)& O(u^2)& O(u^2)\\
O(u^2)& 1+O(u)& O(u^2)\\
O(u^2)& O(u^2)& 1+O(u)
\end{array}
\right)m^2_{SUSY}
\ee
\be
\tilde{m}^2_{eRR}= 
\left(
\begin{array}{ccc}
O(1)& O(t u)& O(t^2 u)\\
O(t u)& O(1)& O(t u)\\
O(t^2 u)& O(t u)& O(1)
\end{array}
\right) m^2_{SUSY}
\ee
\be
\tilde{m}^2_{eRL}= k
\left(
\begin{array}{ccc}
m_e   & c~ m_e u  & c~ m_e u  \\
c~ m_\mu u & m_\mu  &c~ m_\mu u \\
c~ m_\tau u & c~ m_\tau u & m_\tau
\end{array}
\right) m_{SUSY}
\label{RL}
\ee
where $k$ and $c$ are model dependent coefficients.
Depending on the chirality structure of the four-lepton operator we can have different type of suppression:
\bea
{\cal M}( \tau^- \rightarrow \mu^+ e^- e^-)&\propto& \frac{1}{16\pi^2}\frac{m_Z^2}{m_{SUSY}^2}\times u^4~~~~~~~~(LLLL)\nn\\
&\propto& \frac{1}{16\pi^2}\frac{m_Z^2}{m_{SUSY}^2}\times t^3 u^2~~~~~~(RRRR)\nn\\
&\propto& \frac{1}{16\pi^2}\frac{m_Z^2}{m_{SUSY}^2}\times t u^3~~~~~~~(LLRR)
\label{box1}
\eea
\bea
{\cal M}(\tau^- \rightarrow e^+ \mu^- \mu^-)&\propto& \frac{1}{16\pi^2}\frac{m_Z^2}{m_{SUSY}^2}\times u^4~~~~~~~~(LLLL)\nn\\
&\propto& \frac{1}{16\pi^2}\frac{m_Z^2}{m_{SUSY}^2}\times t^2 u^2~~~~~~(RRRR)\nn\\
&\propto& \frac{1}{16\pi^2}\frac{m_Z^2}{m_{SUSY}^2}\times t u^3~~~~~~~(LLRR)
\label{box2}
\eea
The chirality structure (LRLR) is not included because it is strongly suppressed by the ratio $m_i/m_{SUSY}$.
We see that both amplitudes vanish in the limit of exact flavour symmetry, contrary to the expectation based on our effective lagrangian appproach.
In principle the flavour symmetry allows for non-vanishing amplitudes, but the specific SUSY realization considered here prevents
non-vanishing contributions in the exact symmetry limit. Such a result can be justified by analyzing the specific symmetry properties
of the SUSY model. Indeed, it is easy to recognize that, in the limit of exact flavour symmetry lepton masses vanish, and so does the block
$\tilde{m}^2_{eRL}$, whereas $\tilde{m}^2_{eLL}$ is proportional to the unit matrix and $\tilde{m}^2_{eRR}$ is diagonal.
The flavour symmetry in this limit is larger than $A_4\times Z_3\times U(1)_{FN}$: it contains $SU(3)_L\times U(1)_{eR}\times U(1)_{\mu R}\times U(1)_{\tau R}$,
which forbids any flavour-violating transition in the lepton sector. Since both $u$ and $t$ symmetry breaking parameters lies in the percent range,
the predicted rates for $ \tau^- \rightarrow \mu^+ e^- e^-$ and $\tau^- \rightarrow e^+ \mu^- \mu^-$ drop by more than ten order of magnitudes below the present
experimental sensitivity.
Indeed in the SUSY model all the operators of ${\cal L}_{decay}$, eq. (\ref{ldecay}), originate from similar box diagrams and they get the same suppression, thus relaxing also the bounds coming from universality in leptonic muon and tau decays, and from the agreement
between the Fermi constant measured in the muon decay and that extracted from the $m_W/m_Z$ ratio.

Given this strong suppression of the apriori favoured channels $\tau^- \rightarrow \mu^+ e^- e^-$ and $\tau^- \rightarrow e^+ \mu^- \mu^-$, we would like to establish which
is the leading flavour violating process in this SUSY realization of $A_4\times Z_3\times U(1)_{FN}$. We should look for transitions
where a single mass insertion occurs. In ref. \cite{Feruglio:2009iu,Feruglio:2009hu} a detailed analysis of the dipole transitions $\mu \to e \gamma$, $\tau\to \mu \gamma$ and $\tau \to e \gamma$
was presented. In terms of the normalized branching ratios
\be
R_{ij}=\frac{BR(l_i\to l_j\gamma)}{BR(l_i\to l_j\nu_i{\bar \nu_j})}~~~.
\ee
the generic expectation in the SUSY model is
\be
R_{ij}=\frac{6}{\pi}\frac{m_W^4}{m_{SUSY}^4}\alpha_{em}\vert w_{ij} ~u\vert^2
\label{LFV}
\ee
where $\alpha_{em}$ is the fine structure constant and
$w_{ij}$ are dimensionless parameters of order one. Such behavior is due to the dominance of the mass insertion
of $RL$ type, leading to an amplitude linear in the symmetry breaking parameter $u$. As apparent from eq. (\ref{RL})
the amplitude is proportional to the model-dependent coefficient $c$.
There is a class of SUSY model where $c$ actually vanishes \cite{Feruglio:2009iu}. In this case 
the ratios $R_{ij}$ are of the form:
\be
R_{ij}= \frac{6}{\pi}\frac{m_W^4}{m_{SUSY}^4}\alpha_{em}\left[\vert w^{(1)}_{ij} u^2\vert^2+\frac{m_j^2}{m_i^2} \vert w^{(2)}_{ij} u\vert^2\right]
\label{LFVsusy}
\ee
where $w^{(1,2)}_{ij}$ are order one parameters. Even in the case $c=0$ the dipole
transitions appear much more favoured with respect to $\tau^- \rightarrow \mu^+ e^- e^-$ and $\tau^- \rightarrow e^+ \mu^- \mu^-$.
The best present (and future aimed for) limit is for $\mu\to e \gamma$:  
\be
BR(\mu\to e \gamma)<1.2\times 10^{-11}~(10^{-13})
\ee
which implies the following bounds (setting $w_{\mu e}^{(1,2)}\equiv 1$)
\bea
 m_{SUSY}>255~(820)~~{\rm GeV}&&~~~~(u = 0.005)\\
 m_{SUSY}>~0.7~(2.5)~~{\rm TeV}&&~~~~(u = 0.05)~~~.
\eea

\begin{table}[!ht] 
\centering
\begin{math}
\begin{array}{|c|c|}
\hline
&\\[-2pt]
\tt{Process} & \tt{Suppression} \\[2pt]
\hline
\mu^-\to e^- \gamma& t^2 u^2\\
\hline
\tau^-\to \mu^- \gamma& t^2 u^2\\
\hline
\tau^-\to e^- \gamma& u^4\\
\hline
\mu^-\to e^- e^+e^-& t^2 u^2\\
\hline
\tau^-\to \mu^- \mu^+\mu^-& t^2 u^2\\
\hline
\tau^-\to \mu^- e^+e^-& t^2 u^2\\
\hline
\tau^-\to e^- \mu^+\mu^-& u^4\\
\hline
\tau^-\to e^- e^+e^-& u^4\\
\hline
\tau^-\to \mu^+ e^- e^-& u^6 t^2\\
\hline
\tau^-\to e^+ \mu^-\mu^-& u^4 t^4\\
\hline
\end{array}
\end{math} 
\caption{Parametric suppression of the rates for lepton flavour violating processes, assuming the behavior of eq. (\ref{LFVsusy}) for the dipole transitions and of eqs. (\ref{box1}) and (\ref{box2}) for the last two processes.}
\end{table}
\vskip 0.2cm
A related process is the decay $\mu^- \rightarrow e^- e^+ e^-$. Within our SUSY model, at one loop there are contributions both from $\gamma$- and $Z$-penguin diagrams and from box diagrams, 
dominated by a single flavour changing mass insertion in a slepton propagator. The dominant contribution comes from the $\gamma$ penguin diagrams, and we can relate the BR of this process to the 
$\mu \rightarrow e\gamma$ one \cite{para}
\be
\textrm{Br}(\mu^- \rightarrow e^- e^+ e^-) \approx 7 \cdot 10^{-3} \; \textrm{Br}(\mu \rightarrow e\gamma).
\ee
From the current upper bound $Br(\mu^- \rightarrow e^- e^+ e^-) < 1.0 \times 10^{-12}$, using the estimate for $BR(\mu\to e \gamma)$ of eq. (\ref{LFVsusy}), we find
\bea
 m_{SUSY}>140~(225)~~{\rm GeV}&&~~~~ (u = 0.005)\\
 m_{SUSY}>400~(700)~~{\rm GeV}&&~~~~( u = 0.05)~~~,
\eea
where in parenthesis we have shown the results assuming an improvement of the limit  by one order of magnitude.
Other $\tau$ decays such as $\tau^- \to e^-e^+e^-$ ,$\tau^-\to \mu^-\mu^+\mu^-$, $\tau^-\to \mu^- e^+ e^-$, $\tau^- \to e^- \mu^+ \mu^-$,  
have a relationship with $\tau\rightarrow \mu\gamma$ analogous to that between $\mu^- \rightarrow e^- e^- e^+$ and $\mu \rightarrow e\gamma$.
We summarize the suppression of the rates for these processes in Table 2. In this SUSY model there are no LFV transitions that are non-vanishing in the limit of exact flavour symmetry. Flavour violating $\tau$ decays have rates comparable to those of $\mu$ decays,
but the present and future experimental sensitivities are much worse compare to $\mu\to e \gamma$, for which the prospect of detection are the best in this model.

Finally we comment on $\mu$ to $e$ conversion in nuclei.
Although quarks go beyond the description provided by the flavour symmetry $A_4\times Z_3\times U(1)_{FN}$, $\mu$ to $e$ conversion in nuclei can be included in our discussion, since in the SUSY model
considered here it is driven by a $\gamma$ penguin graph similar to that entering the decay $\mu^-\to e^-e^+e^-$. The conversion rate CR($\mu\rightarrow e$ in nuclei) is then related directly to the branching ratio of $\mu\rightarrow e\gamma$, through the relation (the range spans the nuclei used as target in real experiment)
\be
1.5\times10^{-3}\leq \frac{CR(\mu\rightarrow e)}{BR(\mu\rightarrow e\gamma)}\leq 3\times 10^{-2}
\ee
Given the current experimental limit from SINDRUMII \cite{gold_conv} with a gold target, $CR^{Au}\leq 6.1 \times 10^{-13}$ , this channel is not competitive with $\mu\rightarrow e\gamma$. However, if  Mu2e \cite{mu2e} and PRIME \cite{prime} experiments will reach the sensitivity of $CR^{Al} \leq 6\times 10^{-17}$ and $CR^{Ti} \leq 10^{-18}$, they could realistically set the most stringent constraint on lepton flavour violation. For a titanium target $CR^{Ti}=0.5\times 10^{-2}~BR(\mu\rightarrow e\gamma)$ \cite{mue_susy}. Setting $u=0.05$ we get $m_{SUSY}\gtrsim 6.6$ TeV, while for $u=0.005$, $m_{SUSY}\gtrsim 2.3$ TeV.
\section{Conclusion}
Violation of individual lepton numbers have been established in neutrino oscillations 
and might be confirmed in rare transitions involving charged leptons.
Discovering LFV in charged lepton decays would represent a major step towards the solution of the flavour puzzle.
First of all LFV in tau or muon decays at an observable rate requires new physics at an energy scale $M$ not too far from the TeV scale, 
opening the exciting possibility of producing and detecting the new particles responsible for LFV at the LHC.
Moreover, in many models the same parameters that describe neutrino masses and mixing angles are also responsible for LFV,
and testable predictions can be obtained.

Among the proposals that have been formulated in the past to reproduce the nearly tri-bimaximal pattern of lepton mixing angles,
a minimal one, in the sense of group and representations choice, is that based on the $A_4\times Z_3\times U(1)_{FN}$ flavour symmetry.
In this class of models, the spontaneous breaking of the symmetry group is controlled by two small parameters $t$ and $u$ of order few percents, 
and the predictions can be organized as power series in such parameters. For instance deviations from the tri-bimaximal mixing pattern are of order $u$ and $CP$-asymmetries
in right-handed neutrino decays relevant for leptogenesis are of order $u^2$. In order to test this idea it is important to systematically analyze 
the consequences of $A_4\times Z_3\times U(1)_{FN}$ for LFV. In a previous set of papers, radiative decays of charged leptons have been studied,
both in a general effective Lagrangian approach and in a specific SUSY realization of $A_4\times Z_3\times U(1)_{FN}$.
The three decays $\mu\to e \gamma$, $\tau \to \mu\gamma$ and $\tau \to e \gamma$ have similar rates in these models, but
the rates derived from the effective Lagrangian scale as $u^2$ and require rather large values for the scale $M$, while in the SUSY realization
the rates can be much more suppressed and can allow for a more accessible scale of new physics $M$.

In the present work we have extended the analysis to cover all LVF transitions. We have analyzed the most general four-lepton effective Lagrangian
invariant under $A_4\times Z_3\times U(1)_{FN}$. Such a Lagrangian describes several transitions that violate lepton flavour 
and that are not suppressed by any powers of $t$ and/or $u$. All these unsuppressed transitions satisfy the selection rule $\Delta L_e \Delta L_\mu \Delta L_\tau=\pm2$, which 
would provide a nice signature of the assumed flavour symmetry. For instance the decays $ \tau^- \rightarrow \mu^+ e^- e^-$ and $\tau^- \rightarrow e^+ \mu^- \mu^-$
would be favoured over all the other decays with three charged leptons in the final state. Indeed, calling $\varepsilon$ the strength of the 
generic four-lepton interaction in units of the Fermi constant, the non-observation of these tau decays provide the strongest bound on $\varepsilon$ from the existing data:
$|\varepsilon|<(2\div 3)\times  10^{-4}$. If interpreted as a bound on the scale of new physics $M$, this requires $M$ above approximately $10$ TeV. 
Slightly milder bounds comes from the observed universality of leptonic muon and tau decays and from the agreement
between the Fermi constant measured in the muon decay and that extracted from the $m_W/m_Z$ ratio.

In the second part of this paper we have analyzed LFV in the specific SUSY realization of $A_4\times Z_3\times U(1)_{FN}$. In this model, LFV proceeds through one-loop diagrams
where sleptons, charginos and neutralinos are exchanged. Working in the super CKM basis, LFV is due to off-diagonal entries
in the slepton mass matrices and can be analyzed in the mass-insertion approximation.
Contrary to the expectation based on the general effective Lagrangian, a different hierarchy among the rates of LFV transitions is predicted.
Processes where the individual lepton number is violated by a single unit, like $\mu^- \rightarrow e^- e^+ e^-$,
$\tau^- \to e^-e^+e^-$, $\tau^-\to \mu^-\mu^+\mu^-$, $\tau^-\to \mu^- e^+ e^-$, $\tau^- \to e^- \mu^+ \mu^-$
are favoured, with the corresponding rates being suppressed by $u^4\div t^2 u^2$, while
$ \tau^- \rightarrow \mu^+ e^- e^-$ and $\tau^- \rightarrow e^+ \mu^- \mu^-$ are more suppressed,
their rates scaling as $u^6 t^2$ and $u^4 t^4$, respectively. We have traced back this ``anomalous'' behaviour to the fact
that, in the limit of vanishing $u$ and $t$, the low-energy SUSY Lagrangian acquires a much larger flavour symmetry:
$SU(3)_L\times U(1)_{eR}\times U(1)_{\mu R}\times U(1)_{\tau R}$,
which forbids any flavour-violating transition in the lepton sector. We can also understand the predictions of the SUSY case
by inspecting the relevant interaction terms. Once we go in the super CKM basis, neglecting higgsino interactions that
are suppressed by small lepton masses, all the relevant interaction terms are flavour-diagonal. The only source
of flavour violations are the off-diagonal terms of slepton masses. But slepton masses are diagonal when $t$ and $u$ vanish and in this limit
lepton flavour is conserved. The crucial point is that in the SUSY model lepton flavour violation proceeds through bilinear
terms of the low-energy Lagrangian. The symmetry group $A_4\times Z_3\times U(1)_{FN}$ allows for quartic invariant operators 
which violate lepton flavour, such as the effective Lagrangian we have discussed in the first part of this work, but it forbids
any LFV at the level of bilinear terms. Our SUSY model does not represent the most general realization of the 
flavour symmetry $A_4\times Z_3\times U(1)_{FN}$ and leads to more restrictive conclusion about LFV processes.
One the one hand this feature might allow to discriminate the SUSY realization of $A_4\times Z_3\times U(1)_{FN}$ from other
models possessing the same flavour symmetry, but having a more general structure of interaction terms.
On the other hand, as we have seen, in the SUSY model there are several characteristic correlations among LFV transitions,
which could hopefully allow to test $A_4\times Z_3\times U(1)_{FN}$ against other possible underlying flavour symmetries.   
At the moment, given the present experimental sensitivity, $\mu\to e\gamma$ appear as the favorite channel.
In the future, with improved experimental facilities, $\mu$ to $e$ conversion in nuclei could provide the most stringent test of the model.
%
%
\section*{Acknowledgments}
We thank Luca Merlo and Massimo Passera for useful discussions.
We recognize that this work has been partly supported by the European Commission under contract MRTN-CT-2006-035505 and by the European Programme "Unification in the LHC Era", contract PITN-GA-2009-237920 (UNILHC).

\end{document}